\documentclass[10pt,twocolumn,aps,prd]{revtex4-1}
\usepackage{amsfonts,amsmath,amssymb,mathrsfs}
\usepackage{mathtools,mathptmx,array,latexsym}

\def\be{\begin{equation}}
\def\ee{\end{equation}}
\def\bs{\begin{split}}
\def\es{\end{split}}
\def\bea{\begin{eqnarray}}
\def\eea{\end{eqnarray}}
\def\ba{\begin{aligned}}
\def\ea{\end{aligned}}
\def\nn{\nonumber}
\def\cA{\mathcal{A}}

\def\tM{\widetilde{M}}
\def\tV{\widetilde{V}}
\def\tph{\widetilde{\psi_h}}
\def\p{\partial}

\def\sta2{\sin^2\theta}

\begin{document}

\title{Thermodynamical hairs of the four-dimensional Taub-Newman-Unti-Tamburino spacetimes}

\author{Shuang-Qing Wu$^{1}$}
\email{sqwu@cwnu.edu.cn}

\author{Di Wu$^{1,2}$}
\affiliation{$^{1}$College of Physics and Space Science, China West Normal University,
Nanchong, Sichuan 637002, People's Republic of China \\
$^{2}$Department of Physics and Synergetic Innovation Center for Quantum Effect and
Applications, Hunan Normal University, Changsha, Hunan 410081, People's Republic of China}

\date{Received 18 September 2019; Accepted 11 November 2019; Published 26 November 2019}

\begin{abstract}
It is demonstrated that the generic four-dimensional Taub-Newman-Unti-Tamburino (Taub-NUT)
spacetimes can be perfectly described in terms of three or four different kinds of
thermodynamic hairs: the Komar mass ($M = m$), the ``angular momentum" ($J_n = mn$), the
gravitomagnetic charge ($N = n$), and/or the dual (magnetic) mass ($\widetilde{M} = n$). In
other words, the NUT charge is a thermodynamic multihair which means that it simultaneously
has both rotation-like and electromagnetic charge-like characteristics; this is in sharp
contrast with the previous knowledge that it has only one physical feature, or that it is
purely a single solution parameter. To arrive at this novel result, we put forward
a simple, systematic way to investigate the consistent thermodynamic first law and
Bekenstein-Smarr mass formulas of all four-dimensional spacetimes that contain a nonzero NUT
charge, facilitated by first deriving a meaningful Christodoulou-Ruffini-type squared-mass
formula. In this way, not only can the elegant Bekenstein-Hawking one-quarter area-entropy
relation be naturally restored in the Lorentzian and Euclidian sectors of generic Taub-NUT-type
spacetimes without imposing any constraint condition, but also the physical meaning of the
NUT parameter as a poly-facet can be completely clarified in the thermodynamic sense for
the first time.
\end{abstract}


\maketitle

\baselineskip=14pt

\section{Introduction}
Ever since the seminal work of Bekenstein \cite{JDB1973} and Hawking \cite{SWH1976}, it has
been well known that the (Bekenstein-Hawking) entropy of a black hole is proportional to the
area of the horizon and its Hawking temperature to the surface gravity at the horizon. In
terms of the natural unit system, the relations between them are simply given by
\be
S = A/4\, , \qquad T = \kappa/(2\pi) \, .
\label{st}
\ee
The four laws of black hole thermodynamics were also found for asymptotically flat and anti-de
Sitter (AdS) black holes. In particular, the differential first law and the integral
Bekenstein-Smarr mass formula in the $D = 4$ asymptotically flat case read \cite{BCH1973,LS1973}
\be\ba
dM &= T dS +\Omega\, dJ +\Phi\, dQ \,\, +\cdots \, , \\
M &= 2T S +2\Omega\, J +\Phi\, Q \,\, +\cdots \, .
\label{dfl}
\ea\ee
It is remarkable that these mass formulas provide the more elaborate relationship between the
global conserved charges ($M, J, Q$, etc) measured at infinity and the horizon temperature,
entropy, and other quantities ($\Omega, \Phi$) which are evaluated at the horizon but relative
to infinity. When a negative cosmological constant is included for the anti-de Sitter case,
the above mass formulas (\ref{dfl}) should include a modified term $(+V\, dP, -2VP)$,
respectively, where $V$ is the thermodynamic volume conjugate to the pressure $P = 3g^2/(8\pi)$
with $g$ being the inverse of the cosmological radius.

What is more, a Christodoulou-Ruffini-type squared-mass formula was found to be \cite{DC1970,CR1971}
\be
M^2 = \frac{\pi}{4S}\Big(\frac{S}{\pi} +Q^2\Big)^2 +\frac{\pi\, J^2}{S} \,
\label{CRm}
\ee
for the Kerr-Newman black hole, and was later generalized to the Kerr-Newman-AdS$_4$
case \cite{CCK2000}. It should be mentioned that the first law in Eq. (\ref{dfl}) can be
simply deduced \cite{Wu2004} via differentiating the squared-mass formula (\ref{CRm}) with
respect to all of its thermodynamic variables, and then the Bekenstein-Smarr mass formula
can be easily verified.

Recently, the thermodynamics of accelerating (charged and rotating) AdS$_4$ black holes \cite{PD1976}
was discussed in Refs. \cite{AGK2016,AAGKMO2018,AGGKM2019,GS2019} where all of the above formulas
(\ref{st}-\ref{CRm}) were extended to this kind of spacetimes which belong to the class of
double-black-hole solutions with one black hole's event horizon becoming the Rindler horizon.

However, it still seems to be a great exception to asymptotically locally flat spacetimes that
are endowed with a nonzero NUT charge in general relativity. Until now, no consistent thermodynamic
formula similar to the above-mentioned relations (\ref{st}-\ref{CRm}) has been fully and
satisfactorily found for this class of spacetimes in both the Lorentzian and Euclidian sectors,
even for the simplest Taub-NUT spacetime \cite{AHT1951,NTU1963}. As far as the first law (\ref{dfl})
is concerned, to the authors' best knowledge, the differential mass formulas for the NUT-charged
spacetimes that appeared in Refs. \cite{MHA2004,AS2013a,AS2013b,MHA2007,SJ2013,BS2013,UD2016,PP2015,
PP2016} are either inconsistent (here ``inconsistent" means that the thermodynamic quantities cannot
constitute the ordinary conjugate pairs) or even are false. (Nevertheless, it should be noted
that consistent mass formulas for the Demianski-Newman ``black hole" were already conjectured
in Ref. \cite{GPZ2001} without any ``derivation" almost two decades ago!)

In some recent attempts \cite{HKM2019,BGHK2019a,BGHK2019b,BGK2019,RD2019}, the so-called ``consistent
thermodynamical first law" was pursued for the Lorentzian Taub-NUT-type spacetimes. However, these
formulas could not really represent the actual first law from our viewpoint, since the imported
$\bar\psi$-$\mathcal{N}$ pair (which was later called the ``Misner gravitational charge") does
not possess the conventional characteristics of global charges that are measured at infinity;
rather, it combines the contribution of the Misner strings at the horizon, contrary to common wisdom.
Recently, this thermodynamic pair were alternatively explained in Ref. \cite{CG2019} as the angular
velocity and angular momentum of the Misner strings, rather than being interpreted as the temperature
and entropy as in Ref. \cite{BGHK2019a}, since this will seriously challenge the zeroth law of black
hole thermodynamics. According to common sense, it is hardly believable that for a static axisymmetric
Taub-NUT-type spacetime, the temperatures (or surface gravities) at the north and south poles are
different from those at the remaining part of the event horizon. If such an interpretation \cite{BGHK2019a}
is true, then the horizon must be a nonequilibrium system for a stationary black hole, thus violating
the well-established zeroth law.

In this paper, based upon a previously unpublished talk \cite{Wu2015}, we put forward a simple,
systematic routine to investigate the thermodynamics of the four-dimensional spacetimes with a nonzero
NUT charge by first deriving a meaningful Christodoulou-Ruffini-type squared-mass formula, where
a new ``angular momentum" $J_n = mn$ is additionally introduced as an extra conserved charge. Starting
from this squared-mass formula, we derive a consistent first law and Bekenstein-Smarr mass formula
for the NUT-charged spacetimes just like the usual black holes, without assuming that the famous
one-quarter area-entropy relation should hold true in order to get a consistent thermodynamic first
law. In this way, we show that the NUT charge is a thermodynamic multihair, which means that it
simultaneously has both rotation-like and electromagnetic charge-like characteristics. The novelty
of this new viewpoint is that it can plausibly explain many of the peculiar properties of the
NUT-charged spacetime, such as why the NUT parameter has so many different names and why there
are different interpretations of the physical source of Taub-NUT-type spacetimes.

\section{The Lorentzian Taub-NUT geometry}
To begin with, let us first recapitulate some known basic facts of the four-dimensional Taub-NUT
metric in the Lorentzian sector \cite{NTU1963}. We adopt the following line element in which
the Misner strings \cite{CWM1963} are symmetrically distributed along the polar axis:
\bea
&& ds^2 = -\frac{f(r)}{r^2 +n^2}(dt +2n\cos\theta\, d\phi)^2 +\frac{r^2 +n^2}{f(r)}dr^2 \nn \\
&& \qquad~~ +(r^2 +n^2)(d\theta^2 +\sin^2\theta\, d\phi^2) \, ,
\label{nut}
\eea
where $f(r) = r^2 -2mr -n^2$.

The spacetime (\ref{nut}) has many peculiar properties that are mainly due to the presence of wire/line
singularities at the polar axis ($\theta = 0, \pi$), which are often called Misner strings.
Misner \cite{CWM1963} proposed removing these singularities (so as to ensure the metric's regularity)
by introducing a time-periodical identification condition: $\beta = 8\pi\, n$. Then, the unavoidable
presence of closed timelike curves led him \cite{CMW1967} to declare that the NUT parameter is
unphysical and the Taub-NUT spacetime is ``a counter example to almost anything" in general relativity.
The evil consequence is that the dominant community does not usually consider Taub-NUT-type spacetimes
to be black holes (although sometimes they were called ``black holes" in some low-level articles).

In the following, we will derive various mass formulas of four-dimensional Taub-NUT-type spacetimes
without imposing the time-periodicity condition, as was done in Refs. \cite{HKM2019,BGHK2019b,RD2019,CG2019,
WBB1969,AS1971,MR2005,CGG2015}. We will also keep the Misner strings symmetrically
present at the polar axis and only care about the conical singularities where $f(r) = 0$, corresponding
to the outer and inner horizons located at $r_h = r_{\pm} = m \pm\sqrt{m^2 +n^2}$. Below, we will
focus on the (exterior) event horizon; however, the discussions are also true for the interior
(Cauchy) horizon.

The area and surface gravity at the horizon are easily computed via the standard method as
\be
A_h = 4\pi(r_h^2 +n^2)\, , \quad
\kappa = \frac{f^{\prime}(r_h)}{2(r_h^2 +n^2)} = \frac{r_h -m}{r_h^2 +n^2} = \frac{1}{2r_h} \, .
\ee

As for the global conserved charges, the well-known Komar mass at infinity related to the timelike
Killing vector $\p_t$ and the NUT charge or gravitational magnetic (gravitomagnetic) charge
\cite{LBNZ1998} can be computed, respectively, as $M = m$ and $N = n$ \cite{GH2006,CJH1999}.
Its horizon mass \cite{GH2006,MH1997} reads $M_h = r_h -m$. On the other hand,
one can also determine the dual or magnetic-type mass \cite{RS1981,AS1982,DN1966,JSD1974} as
$\tM = n \equiv N$. It is clear that one cannot distinguish the dual or magnetic-type mass from
the gravitomagnetic charge in the present case; however, they will be significantly different
from each other in the case where a nonzero cosmological constant is included.

\section{New charge $J_n = mn$ and squared-mass formula}
In order to derive a reasonable first law, we follow the method used in Ref. \cite{Wu2004} to deduce
a meaningful Christodoulou-Ruffini-type squared-mass formula which is the starting point of our
work. Introducing the ``reduced horizon area" $\cA_h = A_h/(4\pi)$ as in \cite{Wu2004} just
for the sake of simplicity,
\be
\cA_h = r_h^2 +n^2 = 2mr_h +2n^2 \, ,
\label{area}
\ee
and shifting the $2n^2$ term to the left-hand side and squaring the obtained formula, we get the
identity
\be
(\cA_h -2n^2)^2 = 4m^2r_h^2 = 4m^2\cA_h -4m^2n^2 \, .
\label{Sqarea}
\ee

Using only $M = m$ and $N = n$ as the conserved charges would lead to the squared-mass formula
that appeared in Ref. \cite{MR1975}, $4M^2 = (\cA_h -2N^2)^2/(\cA_h -N^2)$, which will give
rise to inconsistent versions \cite{MHA2004,AS2013a,AS2013b,MHA2007,SJ2013,BS2013,UD2016,PP2015,
PP2016} of the first law and integral mass formula. Then, nothing new would take place and the
story would end.

On the contrary, suppose we introduce a new quantity that is closely analogous to the
angular momentum $J = ma$ of the Kerr(-Newman) black hole and is given by
\be
J_n = mn\, .
\ee
Then, we obtain a new squared-mass formula that is almost completely analogous to that of a Kerr-type
black hole presented in Ref. \cite{CR1971},
\be
M^2 = \frac{1}{4\cA_h}(\cA_h -2N^2)^2 +\frac{J_n^2}{\cA_h}
= \frac{\cA_h}{4} +\frac{J_n^2 +N^4}{\cA_h} -N^2\, ,
\label{sqma}
\ee
which forms the basis of our work.

As shown in Ref. \cite{MS2006}, the fact that $J_n = mn \equiv M_5$ corresponds to the mass of a
five-dimensional gravitational magnetic monopole means that it is very natural to consider it as a
conserved charge, at least from the viewpoint of five dimensions. There are also a lot of reasons
to support such an idea. For example, it explains the gyromagnetic ratio \cite{ANA2008,ACD2008}
of Kerr-Taub-NUT-type spacetimes. Furthermore, not only is it related to the gravitational action
\cite{MS2005,AMS2007} of the four-dimensional Taub-NUT-type spacetimes, but it also possesses the
general feature of angular momentum \cite{MR2005,CGG2015,MMR2006}. Recently, it was included in
Ref. \cite{RD2019} to get a consistent first law for the Lorentzian Taub-NUT spacetimes.

It should be pointed out that our identification of $J_n = mn$ as a new ``conserved charge" is the
only input of our procedure. In the following, the complete set of conserved charges for the
Taub-NUT spacetime that we will work with is $M = m$, $N = n$, and $J_n = mn$.

\section{Derivation of differential and integral mass formulas}
Now, as in Ref. \cite{Wu2004}, we can view the mass as an implicit function $M = M(\cA_h, J_n, N)$,
and after differentiating the squared-mass formula (\ref{sqma}) (multiplied by $4\cA_h$) with respect
to its variables we get a new reasonable differential mass formula,
\be
dM = (\kappa/2) d\cA_h +\omega_h\, dJ_n +\psi_h\, dN \, ,
\label{dmf}
\ee
where
\bea
&& \kappa = 2\frac{\p\,M}{\p\,\cA_h}\Big|_{J_n,N} = \frac{\cA_h -2N^2 -2M^2}{2M\cA_h}
= \frac{r_h -m}{r_h^2+n^2} \, , \nn \\
&& \omega_h = \frac{\p\,M}{\p\,J_n}\Big|_{\cA_h,N} = \frac{J_n}{M\cA_h}
= \frac{n}{r_h^2+n^2} \, , \nn \\
&& \psi_h = \frac{\p\,M}{\p\,N}\Big|_{\cA_h,J_n} = \frac{-N(\cA_h -2N^2)}{M\cA_h}
= \frac{-2nr_h}{r_h^2+n^2} \, . \nn
\eea
Then, it is easy to verify directly that the Bekenstein-Smarr integral mass formula is completely
satisfied,
\be
M = \kappa\cA_h +2\omega_h\, J_n +\psi_h\, N \, ,
\label{imf}
\ee
after using the horizon equation $f(r_h) \equiv r_h^2 -2mr_h -n^2 = 0$.

Comparing our new mass formulas presented in Eqs. (\ref{sqma}-\ref{imf}) with the standard ones
(\ref{dfl}-\ref{CRm}), it makes sense to make the familiar identifications
\be
T = \kappa/(2\pi)\, , \qquad S = \pi\cA_h = A/4\, ,
\label{ts}
\ee
which restore the famous Bekenstein-Hawking one-quarter area-entropy relation in a very simple
manner. It is quite remarkable that one should assign a geometric entropy to the Taub-NUT spacetime,
which is just one quarter of its horizon area. In the above ``derivation", we did not require that
the relations (\ref{ts}) should hold true ahead to get a reasonable first law, rather it is a very
natural result via the above thermodynamic deduction.

It is remarkable that, unlike some recent attempts \cite{HKM2019,BGHK2019a,BGHK2019b,BGK2019,RD2019,
CG2019}, our fist law (\ref{dmf}) and the Bekenstein-Smarr mass formula (\ref{imf}) attain their
traditional forms which relate the global conserved charges ($M, N, J_n$) measured at infinity
to those quantities ($T, S, \psi_h, \omega_h$) evaluated at the horizon. At this step, it is quite
reasonable to infer that all four laws of the usual black hole thermodynamics ares applicable to
the Taub-NUT spacetime, which should not longer bear the bad reputation of being ``a counter
example to almost anything". It is now time to formally call it a genuine black hole, at least
as far as its thermodynamics are considered.

\section{Impact of Misner strings}
Without attempting to endow each of the Misner strings with an entropy (and thus also a temperature),
we can see that each string attached at the south and north poles carries the same amount of rotation-like
and electromagnetic-like energies. The total contribution of both strings to the differential and
integral mass formulas (\ref{dmf},\ref{imf}) is a cumulative effect that consists of two terms that
can be rewritten, respectively, as follows:
\bea
&& \omega_h\, dJ_n +\psi_h\, dN = \frac{n^2dm +n(m -2r_h)dn}{r_h^2+n^2} \nn \\
&& \qquad\quad \equiv -\frac{1}{2n}d\Big(\frac{n^3}{r_h}\Big)
 \equiv \frac{1}{n}dJ_n  -\frac{1}{2n} d(nr_h) \, , \\
&& 2\omega_h\, J_n +\psi_h\, N = \frac{2n^2(m -r_h)}{r_h^2+n^2}
 = \frac{-n^2}{r_h} = 2m -r_h \nn \\
&&\qquad\quad \equiv -2\frac{1}{2n}\Big(\frac{n^3}{r_h}\Big)
 \equiv 2\frac{J_n}{n} -2\frac{nr_h}{2n} \, .
\eea
This explains why recent efforts to formulate a ``consistent thermodynamic first law" that is
mathematically consistent were only partially successful \cite{HKM2019,RD2019}. Clearly, neither
the potential $\bar\psi = 1/(4n)$ introduced in Ref. \cite{HKM2019} nor the angular velocity ($1/n$
or $1/(2n)$) proposed in Refs. \cite{RD2019,CG2019} has a well-defined limit when the NUT charge
vanishes. Furthermore, the charges $n^3/r_h$ and $nr_h$ in Refs. \cite{HKM2019,RD2019} do not really have
the general characteristics of a globally conserved charge; rather they must be some quantities related
to the horizon. On the contrary, no such fatal defect exists in the present work, which smoothly
reduces to the Schwarzschild case.

\section{Analytical continuation to the Euclidean sector}
The Euclidean sector \cite{CJH1999,HH1999,HHP1999} is obtained via the Wick rotation $t = i\tau$
and $n = iN$. Now $f(r) = r^2 -2Mr +N^2$, with $M = m$. We will work with the generic values
of the solution parameters ($M, N$), neither introducing the time-periodic identification condition
$\beta = 8\pi\, N$ nor imposing any constraint condition on the solution parameters.

The horizons are located at $r_H = M \pm\sqrt{M^2 -N^2}$, which are determined by $f(r_H) = 0$.
The event horizon area is $A_H = 4\pi\cA_H$, where $\cA_H = r_H^2 -N^2 = 2Mr_H -2N^2$, and its
surface gravity is easily evaluated as
\be
\kappa = \frac{f^{\prime}(r_h)}{2\cA_H } = \frac{r_H -M}{r_H^2 -N^2} = \frac{1}{2r_H} \, .
\label{kp}
\ee

After introducing a conserved quantity $J_N = MN$ as before, we arrive at a squared-mass formula
\be
M^2 = \frac{(\cA_H +2N^2)^2 -4J_N^2}{4\cA_H}\, .
\ee
Similar to the Lorentzian case, one can deduce the following differential and integral mass
formulas:
\be\ba
dM &= (\kappa/2) d\cA_H +\omega_H\, dJ_N +\psi_H\, dN \, , \\
M &= \kappa\cA_H +2\omega_H\, J_N +\psi_H\, N \, ,
\ea\ee
where $\kappa = (r_H -M)/\cA_H$, $\psi_H = 2Nr_H/\cA_H$, and $\omega_H = -N/\cA_H$.

It is natural to suggest that one should identify a geometric entropy with $S = A_H/4 = \pi\cA_H$
and the Hawking temperature via $T = \kappa/(2\pi)$, so that a fairly satisfactory relation also holds
true for the thermodynamics of the Euclidean Taub-NUT solution with the generic parameters $(M, N)$,
which can be roughly interpreted as a spinning Misner string with angular momentum $J_N = MN$ and
the gravitomagnetic charge $N$.

Now, let us impose the periodic condition $\beta = 8\pi N = 2\pi/\kappa$ and discuss the nut and
bolt cases, separately. A self-dual Taub-NUT solution is the special case when $M = N$, with
a nut at $r_H = N$ with zero horizon area ($\cA_H = 0$) and surface gravity $\kappa = \p_r[(r-N)/(r+N)]
|_{r=N}/2 = 1/(4N)$. In this case, both $\omega_H$ and $\psi_H$ become infinite; however, the
compositions
\be\ba
 \omega_H\, dJ_N +\psi_H\, dN = dN \equiv & dM \, , \\
 2\omega_H\, J_N +\psi_H\, N = N \equiv & M \,
\ea\ee
remain finite. [For example, one can let $r_H = N +\epsilon$ and then take the $\epsilon \to 0$
limit. Note that when this limit is applied to Eq. (\ref{kp}), an additional factor of $1/2$ should
be multiplied to get the correct value $\kappa = 1/(4N)$ for the surface gravity.] The contribution
of the Misner strings to the nut solution is $N^2/r_H = N$.

In contrast, the regular bolt solution is obtained when $M = 5N/4$. In this case, the horizon is
located at $r_H = 2N$, and the other quantities are $\kappa = 1/(4N)$, $\cA_H = 3N^2$, $\omega_H
= -1/(3N)$, $J_N = 5N^2/4$ and $\psi_H = 4/3$, so it is easy to show that
\be\ba
\omega_H\, dJ_N +\psi_H\, dN = dN/2 \equiv & M -(\kappa/2)\,d\cA_H \, , \\
2\omega_H\, J_N +\psi_H\, N = N/2 \equiv & M -\kappa\cA_H \, ,
\ea\ee
which shows that the contribution of the Misner strings to the bolt solution is $N^2/r_H = N/2$.

Since the bolt and nut solutions are matched with the identical surface gravity $\kappa = 1/(4N)$,
their mass difference is $\Delta\, M = 5N/4 -N = N/4$, while their reduced area difference is
$\Delta\,\cA_H = 3\pi N^2$. Thus, the contribution of the Misner strings is $\Delta\, M
-\kappa\Delta\, \cA_H = N/4 -3N/4 = -N/2$, which coincides with the total contribution of the Misner
strings $N/2 -N = -N/2$.

One can understand that the entropy obtained by Hawking \emph{et al} \cite{CJH1999,HH1999,HHP1999}
and Mann \cite{RBM1999} is a (generalized) relative entropy, which is computed for the bolt solution
with respect to the reference background, namely, the self-dual nut solution.

\section{Adding a nonzero negative cosmological constant and electric charge}
Now we can extend the above work to the Lorentzian Reissner-Nordstr\"{o}m-Taub-NUT-AdS$_4$ spacetime
with a nonzero cosmological constant and a pure electric charge. The metric is still given by Eq.
(\ref{nut}) with the parameters $(m, n)$ replaced by $(M, N)$, and now $f(r) = r^2 -2Mr -N^2 +Q^2
+g^2(r^4 +6N^2r^2 -3N^4)$. In addition, the electromagnetic gauge potential one-form is
\be
\mathbf{A} = \frac{Qr}{r^2+N^2}(dt +2N\cos\theta\, d\phi) \, .
\ee
It is easy to compute the electric charge $Q$ and gravitomagnetic charge $N$ of the spacetime,
while its electric mass $M$ and dual (magnetic) mass \cite{AAMO2016} $\tM = N(1+4g^2N^2)$, which
is different from the NUT charge now, can be calculated using the conformal completion method.

The solution admits Killing horizons which are determined by $f(r_h) = 0$, where the electrostatic
potential is $\Phi = Qr_h/(r_h^2+N^2)$. The horizon area is $A_h = 4\pi\cA_h$ and the surface gravity
is given by
\be
\kappa = \frac{f^{\prime}(r_h)}{2\cA_h} = \frac{r_h -M +2g^2(r_h^2 +3N^2)r_h}{\cA_h} \, ,
\ee
where $\cA_h = r_h^2 +N^2 = 2Mr_h +2N^2 -Q^2 -g^2(r_h^4 +6N^2r_h^2 -3N^4) = 2Mr_h +2N^2 -Q^2
-g^2(\cA_h^2 +4N^2\cA_h -8N^4)$.

Now, after introducing $J_N = MN$, squaring the identity $2Mr_h = (1+4g^2N^2)(\cA_h -2N^2) +Q^2
+g^2\cA_h^2$, and adding a term $4M^2N^2$ to it, we get
\bea
M^2\cA_h = J_N^2 +\frac{1}{4}\big[(1+4g^2N^2)(\cA_h -2N^2) +g^2\cA_h^2 +Q^2\big]^2 \, , \nn
\eea
which is nothing but the squared-mass formula
\bea
M^2 &=& \frac{1}{\cA_h}\Big[\Big(1 +\frac{32\pi}{3}P\, N^2\Big)(\cA_h -2N^2) +Q^2 \nn \\
 && +\frac{8\pi}{3} P\, \cA_h^2\Big]^2+\frac{J_N^2}{\cA_h}\, ,
\label{Msq}
\eea
where $P = 3g^2/(8\pi)$ is the generalized pressure.

The differentiation of the mass formula (\ref{Msq}) leads to the first law
\be
dM = (\kappa/2) d\cA_h +\omega_h\, dJ_N +\psi_h\, dN +\Phi\, dQ +V\, dP \, ,
\ee
where
\bea
&&\kappa = \frac{r_h -M +2g^2(r_h +3N^2)r_h}{\cA_h} \, , \quad \Phi = \frac{Qr_h}{\cA_h} \, , \nn \\
&& \psi_h = 2Nr_h\frac{-1 +2g^2(r_h^2 -3N^2)}{\cA_h} \, , \quad \omega_h = \frac{N}{\cA_h} \, , \nn \\
&& V = \frac{4\pi(r_h^4 +6N^2r_h^2 -3N^4)r_h}{3\cA_h} \, . \nn
\eea
Then we can directly verify that the Bekenstein-Smarr mass formula
\be
\kappa\cA_h +2\omega_h\, J_n +\psi_h\, N +\Phi\, Q -2VP = M \,
\ee
is completely satisfied. It is natural to recognize $S = A_h/4 = \pi\cA_h$ and $T = \kappa/(2\pi)$, so
that the solution behaves like a genuine black hole without violating the beautiful one-quarter
area/entropy law. In sharp contrast with Refs. \cite{HKM2019,BGHK2019a,BGHK2019b,BGK2019,RD2019}, here we
neither insist that this law be obeyed nor require that the first law and the integral mass formula
be consistent. This is a very natural product of the pure thermodynamic deduction.

In the above, the derived conjugate thermodynamic volume $V$ is not equal to $\tV = 4\pi(r_h^2
+3N^2)/3$ as given in Refs. \cite{HKM2019,BGHK2019a,BGHK2019b,BGK2019}. If one prefers to use such a
thermodynamic volume, then the dual (magnetic) mass $\tM = N(1+4g^2N^2)$ can be further introduced
as an additional conserved charge into the differential and integral mass formulas:
\bea
dM &=& (\kappa/2) d\cA_h +\omega_h\, dJ_N +\tph\, dN +\zeta\, d\tM
 +\Phi\, dQ +\tV\, dP \, , \nn \\
M &=& \kappa\cA_h +2\omega_h\, J_n +\tph\, N +\zeta\tM +\Phi\, Q -2\tV\, P \, , \nn
\eea
where
\bea
\tph = -\frac{2Nr_h}{\cA_h} -(1 -4g^2N^2)\zeta\, , \quad
\zeta = \frac{r_h(r_h^2-3N^2)}{4N\cA_h} \, , \nn
\eea
suggesting that the NUT charge is a thermodynamic trihair rather than a bihair. We will not
attempt to discuss the thermodynamics of its Euclidean sector further \cite{CEJM1999,EJM1999}.

\section{Extension to the Kerr-Newman-Taub-NUT spacetime}
Finally, let us discuss the general case including a nonzero rotation parameter but without a cosmological
constant. The line element of the Kerr-Newman-Taub-NUT \cite{JGM1973} or Demianski-Newman \cite{DN1966}
spacetime with the Misner strings symmetrically distributed along the rotation axis and the electromagnetic
one-form are
\bea
ds^2 &=& -\frac{\Delta(r)}{\Sigma}\big[dt +(2N\cos\theta -a\sta2)d\phi\big]^2
 +\frac{\Sigma}{\Delta(r)}dr^2 \nn \\
&& +\Sigma d\theta^2 +\frac{\sta2}{\Sigma}\big[a dt -(r^2+a^2+N^2)d\phi\big]^2 \, , \\
\mathbf{A} &=& \frac{Qr}{\Sigma}\big[dt +(2N\cos\theta -a\sta2)d\phi\big] \, ,
\eea
where $\Sigma = r^2 +(N +a\cos\theta)^2$ and $\Delta(r) = r^2 +a^2 -2Mr -N^2 +Q^2$.

The global conserved charges for this spacetime are the Komar mass $M$, angular momentum
$J = Ma$, electric charge $Q$, and gravitomagnetic charge or dual (magnetic) mass (both
of which are identical to the NUT charge $N$).

The horizons are determined by $\Delta(r_h) = 0$, which gives $r_h = M \pm\sqrt{M^2 +N^2
-Q^2 -a^2}$. The event horizon area is $A_h = 4\pi\cA_h$, where $\cA_h = r_h^2 +a^2 +N^2 =
2Mr_h +2N^2 -Q^2$. At the horizon, the surface gravity, angular velocity, and electrostatic
potential can be evaluated via the standard method as
\be
\kappa = \frac{\Delta^{\prime}(r_h)}{2\cA_h} = \frac{r_h -M}{\cA_h} \, , \quad
\Omega = \frac{a}{\cA_h} \, , \quad \Phi = \frac{Qr_h}{\cA_h} \, .
\label{kOP}
\ee

Following the above procedure, we square the identity $2Mr_h = \cA_h -2N^2 +Q^2$, and
then after adding $4M^2(a^2 +N^2)$ to it, we can obtain the useful identity
\bea
4M^2\cA_h = 4J^2 +4M^2N^2 +(\cA_h -2N^2 +Q^2)^2 \, , \nn
\eea
which is exactly our squared-mass formula
\be
M^2 = \frac{(\cA_h -2N^2 +Q^2)^2 +4J_N^2 +4J^2}{4\cA_h} \,
\label{MaSq}
\ee
if we introduce $J_N = MN$ as a new conserved charge, as before.

Differentiation of the above squared-mass formula (\ref{MaSq}) yields the first law
\bea
dM = (\kappa/2) d\cA_h +\Omega\, dJ +\omega_h\, dJ_N +\psi_h\, dN +\Phi\, dQ \, ,
\eea
where $(\kappa, \Omega, \Phi)$ are given by Eq. (\ref{kOP}) and
\be
\omega_h = \frac{N}{\cA_h} \, , \quad \psi_h = -\frac{2Nr_h}{\cA_h} \, .
\ee
One can verify that the integral mass formula
\be
\kappa\cA_h +2\Omega\, J +2\omega_h\, J_n +\psi_h\, N +\Phi\, Q = M \,
\ee
is completely satisfied. This completes the simple algebraic derivation of the mass formulas
conjectured in Ref. \cite{GPZ2001} for the Demianski-Newman ``black hole".

The consistency of the above thermodynamic formalism suggests that one should restore the
well-known Bekenstein-Hawking area/entropy relation $S = A_h/4 = \pi\cA_h$ and Hawking
temperature $T = \kappa/(2\pi)$, which means that the whole class of NUT-charged spacetimes
should be viewed as generic black holes.

\section{Concluding remarks}
In this work we have presented a simple, systematic way to naturally derive the thermodynamical
first law and Bekenstein-Smarr mass formula of four-dimensional Taub-NUT-type spacetimes. This
might be the most appropriate candidate framework to address the longstanding problem of the
thermodynamics of both Lorentizan and Euclidean Taub-NUT-type spacetimes with the generic
parameters. Not only can the beautiful Bekenstein-Hawking one-quarter area-entropy relation be
naturally restored, but also all four laws of the usual black hole thermodynamics are
shown to be completely applicable to the Taub-NUT-type spacetimes, without imposing any constraint
condition. Furthermore, the physical meaning of the NUT parameter as a multihair has been
clarified via its thermodynamics, namely, it can explain for the first time why the NUT parameter
has so many different names and why there are different interpretations of the physical source of
Taub-NUT-type spacetimes.

It would be suitable to think that our mass formulas properly describe a thermodynamic system made
up of the horizon and two turning points (soliton and antisoliton pair sitting at two south and north
poles) rather than all of theMisner strings, whose impacts can be coherently decomposed into two parts:
rotation-like and electromagnetic charge-like effects.

\section*{Acknowledgments}
This work is partially supported by the National Natural Science Foundation of China under Grants No.
11675130, No. 11275157, No. 10975058 and No. 11775077. The main content of this work was presented \cite{Wu2015} at the workshop held at
NingBo University, November 26, 2015. The latter was based upon the first author's unpublished draft
(July 18, 2006).


\end{document}